\documentclass[aps,pra,10pt,amsmath,amssymb,twocolumn,superscriptaddress,showpacs]{revtex4-1}
\usepackage{graphicx}
\usepackage[caption=false]{subfig}
\usepackage{dcolumn}
\usepackage{bm}
\usepackage{natbib}

\begin{document}

\title{Polarization Self-rotation in Ultracold Atomic $^{87}$Rb}

\author{Travis Horrom}
\affiliation{Department of Physics, The College of William and Mary, Williamsburg, VA 23187, USA}

\author{Salim Balik}
\affiliation{Department of Physics, Old Dominion University, Norfolk, VA 23529, USA}

\author{Arturo Lezama}
\affiliation{Instituto de F\'isica, Universidad de la Rep\'ublica, Casilla de correo 30, 11000, Montevideo, Uruguay}

\author{Mark D. Havey}
\affiliation{Department of Physics, Old Dominion University, Norfolk, VA 23529, USA}

\author{Eugeniy E. Mikhailov}
\affiliation{Department of Physics, The College of William and Mary, Williamsburg, VA 23187, USA}

\date{\today}

\begin{abstract}
We report on a combined experimental and theoretical study of polarization self-rotation in an ultracold atomic sample.  In the experiments, a probe laser is tuned in the spectral vicinity of the D$_1$ line to observe polarization self-rotation in a sample of ultracold $^{87}$Rb prepared in a magneto-optical trap.  Systematic measurements of the rotation angle of the light-polarization ellipse as a function of laser intensity, initial ellipticity and detuning are made.  The observations, in good agreement with theoretical simulations, are indicative of the presence of a residual static magnetic field, resulting in measured asymmetries in the rotation angle for right and left ellipticities.  In this paper we present our detailed experimental results and analysis of the combined influences of polarization self-rotation and the Faraday effect.
\end{abstract}

\maketitle

\section{Introduction}
It is well established that the polarization ellipse of light can rotate when elliptically polarized light interacts with near-resonant atoms.  This nonlinear effect, termed polarization self-rotation (PSR), has been studied extensively both theoretically and experimentally for hot atomic vapor
samples~\cite{RevModPhys.74.1153, PhysRevA.63.043814, Davis:92, PhysRevA.66.043815, PhysRev.137.A801, 2002JMOp...49.2565N}. Unlike the Faraday effect, PSR does not require the presence of an external magnetic field. Its distinctive feature is the dependence on light ellipticity. Right or left handed ellipticities result in an opposite sense of rotation and the rotation angle is proportional to the ellipticity. In the presence of a magnetic field PSR is superimposed to other effects.  Of principal importance here is Faraday rotation resulting in the loss of symmetry in the response of the atomic gas to light of  right and left elliptical polarizations. To date, very few studies have been concerned with PSR in the presence of a magnetic field. Though simple PSR is observed with no external magnetic field, the presence of such a field will change the rotation due to added shifts to the Zeeman substates, thus resulting in Faraday rotation.

As a practical matter, a good understanding of the overall rotation effect could be important in experiments using a cold atom trap, where the magnetic field is characteristically  not negligible in the entire sample region.  In particular, self-rotation, shown to be sensitive to magnetic fields, could be a useful diagnostic tool for characterizing the local magnetic field environment. Furthermore, in many experiments there are transient magnetic fields in the sample environment generated through eddy currents arising from switching the trap magnetic fields.   While PSR has been studied in hot atomic vapors, it has not, to our knowledge, been studied extensively in a Doppler-free cold atomic sample. There have been studies carried out in cold atom traps studying scattering and Faraday rotation effects~\cite[e.g.]{PhysRevA.79.033418, PhysRevA.64.033402, 2003APS..DMP.L3010N, 2010PhRvA..81e3420W}, and a study of PSR will complement this work while leading to a better understanding of cold atom dynamics.

Of more general interest, a promising and fundamental motivation for study of the PSR effect in ultracold atoms comes from the
possible application of these systems to the generation of squeezed light.
PSR is known to be a mechanism that leads to squeezed states of light where
quantum noise fluctuations drop below the standard quantum limit (SQL). The
relationship between self-rotation and squeezing was analyzed in detail by Matsko et al.~\cite{PhysRevA.66.043815}.  The wide range of potential applications for a source of highly squeezed light include communications, precision measurements, and quantum information.  In quantum information implementations, for example, light storage experiments can use squeezed quantum states to test the efficiency of optical quantum information storage.  Polarization self-rotation in hot Rubidium vapors has
been shown experimentally to lead to vacuum squeezing with noise suppression on the order of 1 dB below the standard quantum limit \cite{Mikhailov:08,2009JMOp...56.1985M, PhysRevA.68.025801, Agha:10}.  It has been suggested in~\cite{2009JMOp...56.1985M} and in~\cite{PhysRevA.73.023806} that a higher level of quadrature squeezing may be seen in a cold atom sample by taking advantage of the nearly stationary atoms and associated negligible Doppler broadening.  While other effective methods of squeezing have been demonstrated using nonlinear crystals and fibers for example \cite{PhysRevLett.100.033602, Siz99}, these methods are often limited to specific wavelengths, and so a source of squeezed light at a frequencies near atomic transitions is desirable for many applications where atomic samples are used for storage and processing of quantum information.

In this paper, we report on a systematic study of polarization
self-rotation in an ultracold atom medium.  In particular, we focus our
attention on the differences between the PSR effect observed in atomic
$^{87}$Rb samples contained either in a magnetically shielded vapor cell or
in a magneto optical trap, where the atoms explore regions of non-zero
magnetic field.  This study of PSR focuses on a quantitatively different
sample type, and also explores implementation of PSR in the presence of the
local magnetic field.   As we will see, the presence of a relatively small
magnetic field has profound effects on the observed ellipticity rotation in
the case of ultracold samples.  In the following sections, we first provide
a brief review of the fundamentals of polarization self-rotation. This is
followed by an overview of the experimental arrangement, and those features
of particular importance to the studies discussed here.  A sketch of the
model we use to quantitatively examine the results is followed by a presentation and discussion of the experimental measurements.  We close with an overview and perspectives on the results.

\section{Polarization Self Rotation}
The polarization state of a classical monochromatic beam of light may be described in terms of two circularly polarized
components $\sigma ^+$ and  $\sigma ^-$.  For linearly or elliptically
polarized light, these two components have a stable relative phase.
The amplitudes of the $\sigma ^+$ and  $\sigma ^-$ components are
equal for linearly polarized light and unequal for an elliptically
polarized beam.
If the light interacts
with a near-resonant atomic transition, the imbalance of the intensities of the two circular polarization components generally results in unequal coupling with the different Zeeman substates, the details of which depend on the specific transitions of interest. This leads to differences in the light-shifts and the populations, via optical pumping, of Zeeman substates with magnetic quantum number $m$ of opposite sign. As a result, the refractive index of the effective medium is different for the two circular components; this results in rotation of the polarization ellipse.  The self-rotation angle $\theta$ is given by $\theta = g\varepsilon (0)L$ where $\varepsilon (0)$ is the incident small light ellipticity, $L$ is the length of propagation, and $g$ is a self-rotation parameter dependent on the atomic medium as well as the laser intensity and frequency~\cite{PhysRevA.66.043815}. At low intensities $g$ is linearly dependent on the light intensity. For an isolated atomic transition, if $\varepsilon$ and $L$ are held constant, the self-rotation angle presents an
antisymmetric dispersive shape as a function of the spectral detuning with respect to the unperturbed atomic transition. In multilevel systems, off resonance transitions associated with the presence of nearby states may distort the symmetry of the PSR response around a given transition (see the discussion in Rochester $\emph{et al.}$ ~ \cite{PhysRevA.63.043814}).  As both the intensity of the probe field and the initial ellipticity change the strengths of the $\sigma ^+$ and  $\sigma ^-$ polarization components, the self-rotation angle due to PSR will be proportional to these two factors.  However, this assumes that there is no external magnetic field influencing the atoms. It is well known that the presence of an applied magnetic field will also cause circular birefringence leading to polarization rotation due to the Faraday effect.  In an atomic sample with a small external magnetic field, the observed rotation will depend on both mechanisms, and it is the interplay of these two mechanisms that mainly concerns us here.

\begin{figure}
\includegraphics[width=\columnwidth]{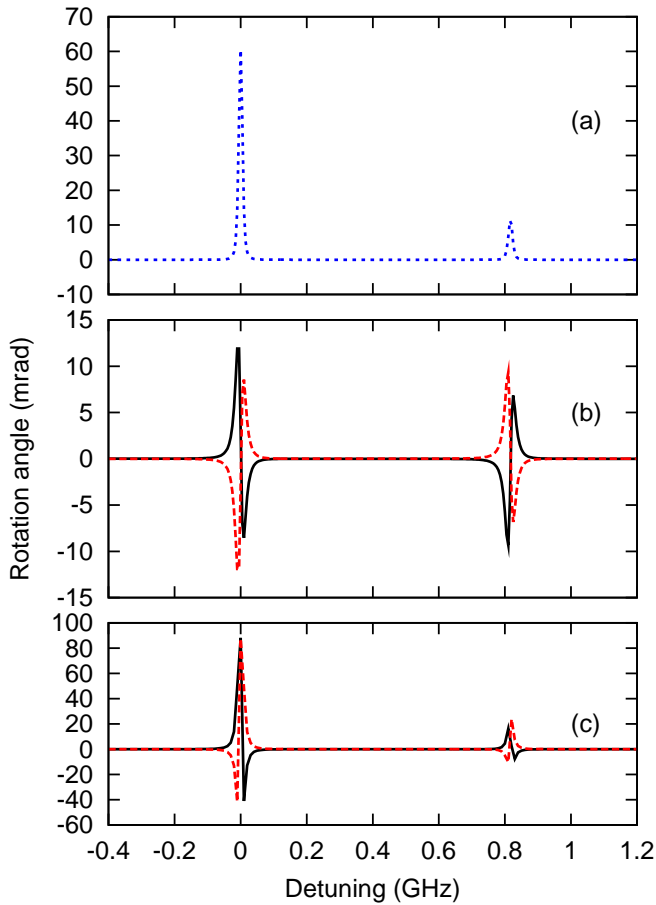}
\caption{Color online. Calculated polarization rotation around the $F_g=2
\rightarrow F_e=1$ (corresponds to zero detuning) and $F_g=2 \rightarrow
F_e=2$ hyperfine transitions as a function of detuning. (a) Pure Faraday
rotation ($ B=0.01\Gamma, \varepsilon =0$). (b) Pure PSR rotation ($B=0,
\varepsilon =\pm 25^{\circ}$),  black solid (red dashed) lines correspond
to positive (negative) ellipticity.
(c) Combined Faraday and PSR effects ($ B=0.01\Gamma, \varepsilon =\pm
25^{\circ}$). Parameters: $C=3$, $I=2$~mW/cm$^{2}$, $\gamma=0.001\Gamma$ \label{Bfield effect}}
\end{figure}

The influence of the two mechanisms is illustrated in Fig.~\ref{Bfield effect} showing a numerical simulation (presented in
Section~\ref{NumSim}) of the PSR effect for the $F_g=2 \rightarrow F_e=1$
hyperfine transition. We note here that in this figure, and elsewhere in this paper, the spectral location of this hyperfine transitions of the D$_1$ line of $^{87}$Rb corresponds to the zero of spectral detuning, and the $F_g=2 \rightarrow F_e=2$ then corresponds to a positive 815~MHz detuning. Fig.~\ref{Bfield effect}a shows the rotation due to the Faraday effect alone, which is due to the presence of an applied magnetic field. The angle of rotation has a fixed sign determined by the orientation of the magnetic field. Notice the difference in magnitude of the Faraday effect for the two hyperfine transitions. Fig.~\ref{Bfield effect}b shows the PSR effect alone (no magnetic field) for two opposite incident field ellipticities ($\pm 25^{\circ}$). As expected, the resonances have dispersion-like shapes with small asymmetries due to the neighboring transition. The decrease of the rotation angle with increasing detuning from resonance is considerably slower than for the Faraday effect. Interestingly enough, the magnitude of the PSR effect is quite similar for the two hyperfine transitions although with opposite signs. Fig.~\ref{Bfield effect}c shows the combined effect of the two mechanisms. The Faraday effect is responsible for the lack of symmetry for opposite ellipticities and for the imbalance between the two transitions. In all cases, the PSR effect is dominant for larger detunings.

\section{Experimental Arrangement}

\begin{figure}
\includegraphics[width=\columnwidth]{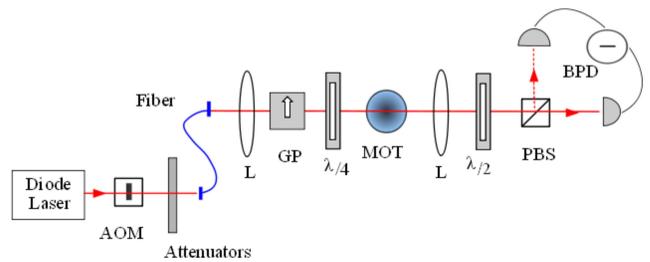}
\caption{Color online. Schematic diagram of the experimental arrangement\label{experiment setup}.}
\end{figure}

\begin{figure}
\includegraphics[width=.8\columnwidth]{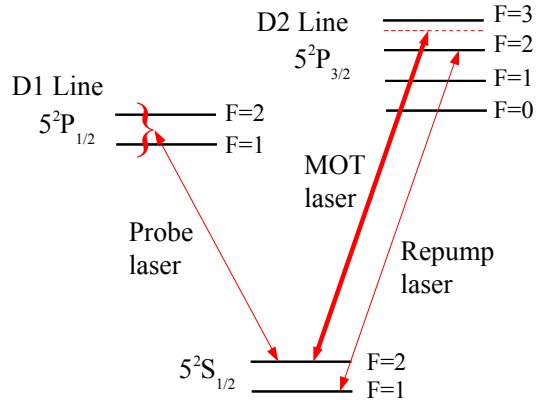}%
\caption{Color online. Partial diagram of the $^{87}$Rb levels scheme\label{level scheme} indicating the trapping and probe transitions.}
\end{figure}

A schematic diagram of the experimental arrangement is shown in Fig.~\ref{experiment setup}.  The
rubidium atom trap is set up in a standard six beam magneto optical
trapping configuration.  This trap, and associated diagnostics, has been
described in detail in \cite{PhysRevA.79.033418}.  In the arrangement,
light from an external cavity diode laser delivers a total power of
$\approx$ 20~mW  to the atom sample. The laser is spectrally detuned 18~MHz
below the $F_g=2 \rightarrow F_e=3$ $^{87}$Rb D$_2$ hyperfine transition.
A weaker repumper laser having a power of $\approx$ 3~mW is tuned to
resonance with the $F_g=1 \rightarrow F_e=2$ D$_2$ transition, thus
maintaining most of the atomic population in the $5^2 S_{1/2}$, F=2 ground
state.  A level scheme with the different laser frequencies used is shown
in Fig.~\ref{level scheme}.  Absorption imaging of the sample shows that it
contains about $7\times 10^7$ $^{87}$Rb atoms.  Ballistic expansion
measurements give a typical temperature of 300 $\mu K$ for the atom sample.
The sample is well described as a sphere with a spatially Gaussian atom distribution
having a Gaussian radius of about 500~$\mu$m. The sample has a
peak density of about $7\times 10^9$ atoms/cm$^3$ and an optical depth on
the order of 2 for the transitions of this study. The trap magnetic field
gradient is variable, with a typical value of 5 G/cm. Application of the
MOT and repumper lasers to the sample is manipulated by computer switched
acousto optical modulators. In most measurements, the trapping beams were
turned off while the probe was on. The repumping laser and the trap
magnetic field were left on continuously. Turning the trapping beams off
during the measurement results in the expansion of the atomic cloud, with
its radius growing at an approximate ballistic rate of 200 $\frac{\mu
\text{m}}{\text{ms}}$.

An external cavity diode laser tuned to the $^{87}$Rb D$_1$ line ($\lambda
\sim 795$ nm) serves as the probe beam.  The probe frequency is scanned
across the $F_g=2 \rightarrow F_e=1$ and $F_g=2 \rightarrow F_e=2$
hyperfine transitions and monitored with a wavemeter.  An acousto-optic
modulator (AOM) is used for fast switching of the probe beam, while neutral
density filters are used for power attenuation.  The light beam is launched
into a single-mode optical fiber to achieve a high quality and nearly
Gaussian output beam intensity distribution.  The fiber output passes
through a high quality Glan polarizer (GP) to ensure linear polarization
and a quarter-wave plate to control the ellipticity. The beam is focused
into the cold-atom cloud with a beam diameter around 250 $\mu$m ($1/e^2$) in the interaction region. Maximum available probe beam power is $\approx$ 2~mW.

Once the probe laser exits the MOT chamber, a half-wave plate sets the
probe beam light polarization angle to $45^\circ$ with respect to a
polarizing beam splitter (PBS) which then separates two orthogonal
components. These two beams are then directed to a custom built balanced
photodetector (BPD) where the signals from the two polarization components
are subtracted.  The BPD includes two matched Hamamatsu S5106 photodiodes,
with quantum efficiency 93\%, and a low-noise high-bandwidth TI OPA842
operational amplifier.  Rotating the quarter-wave plate before the MOT to
control the ellipticity also changes the angle of the major polarization
axis of the beam; the half-wave plate is adjusted to bring this angle back
to $45^{\circ}$  by zeroing the balanced signal in the absence of the atom
sample.  As one consequence of this, any imbalance of the orthogonal
polarization components is due to rotation of the  polarization ellipse caused by the atoms. The rotation angle is proportional to the subtracted signal according the expression:
\begin{equation}
	\tfrac{I_1-I_2}{I_1+I_2}=\sin^2(\tfrac{\pi}{4} + \theta) - \cos^2(\tfrac{\pi}{4}+\theta) = \sin(2\theta)\simeq 2\theta
\end{equation}
where $I_1$ and $I_2$ are the intensities incident upon the two photodetectors and $\theta$ is the PSR angle in radians.

\section{Overview of Numerical Simulations}
\label{NumSim}
To compare the simulations with the experimental results, we performed a numerical
calculation of the PSR angle for parameters approaching the experimental
conditions. However, unlike in the experiment, the calculation is carried
out for a homogeneous atomic sample of motionless atoms. As a consequence
of this limitation in our modeling, the presence of the spatially
inhomogeneous magnetic field, the MOT cloud expansion during the
measurement phase, and the acceleration of atoms due to light forces are
not directly taken into account.  The PSR angle was numerically calculated
by solving the optical Bloch equations for the atomic system in the
presence of an elliptically polarized monochromatic classical light field
with ellipticity $\varepsilon$. The calculation is similar to the one
presented in \cite{lezama02}. We briefly outline the main ingredients. We
consider light propagation along axis $z$ through an homogeneous atomic
sample in the presence of a constant magnetic field. The major axis of the
incident light polarization
ellipsis is taken along $x$. Taking into account the level structure of the D$_1$ transition of the $^{87}$Rb atom, we include in the calculation a single ground state hyperfine level with total angular momentum $F_{g}=2$ and two excited hyperfine levels with angular momenta $F_{e}=1$ and $F_{e}'=2$. All Zeeman substates are taken into account for these 3 levels. The decay of the excited states is due to spontaneous emission at a rate $\Gamma$. In addition, the transit time decay is accounted for by an overall decay rate parameter $\gamma$ ($\gamma \ll \Gamma$). The magnetic field strength $B$ is measured in units of the corresponding Zeeman frequency shift. The incident field with electric field amplitude $E$ has a Rabi frequency $\Omega = \mu E/ \hbar$  where  $\mu$ is the \emph{reduced} dipole moment matrix element for the $5S_{1/2}\rightarrow 5P_{1/2}$ D$_1$ transition. In our modeling, the atomic medium has an optical depth $4C$  where $C\equiv\frac{\eta L \omega \mu^{2}}{2 \varepsilon_{0} \Gamma c \hbar}$ is the cooperativity parameter ($\eta$ is the atomic density, $L$ the medium length). Writing the electric dipole operator as $\mathbf{D}=\mu \mathbf{S}$, the matrix elements of the dimensionless operator $\mathbf{S}$ for the different transitions between Zeeman substates are evaluated using standard angular momentum algebra \cite{Rb87numbers}. We numerically solve the Bloch equations for the steady state normalized density matrix $\rho$. The polarization ellipse rotation angle $\theta$ is given by the accumulated phase difference between the two circular components of the light:
\begin{equation*}\label{caculated angle}
	\theta=\frac{\sqrt{2}C\Gamma}{\Omega}
	\text{Real}\left\{
	\text{Tr}\left[
	\rho \mathbf{S}\cdot
	\left(
	\frac{\mathbf{e_{+}}}{\cos\varepsilon+\sin\varepsilon}-\frac{\mathbf{e_{-}}}{\cos\varepsilon-\sin\varepsilon}
	\right)
	\right]
	\right\}
\end{equation*}
where $\mathbf{e_{+}}$ and $\mathbf{e_{-}}$ are complex unit vectors corresponding to the two circular polarizations.

\section{Experimental Results}
In the experiments, we investigated the angle of rotation due to PSR under
a variety of different conditions to explore as large as possible a zone of experimental parameter space.  We first sought the best conditions of the MOT for the largest obtainable self-rotation (Section~\ref{preparation}).  This included the presence or absence of the trapping laser beam, the repumping beam, and the MOT gradient magnetic field.  We then varied the initial ellipticity of the probe laser to confirm the presence of polarization self-rotation (Section~\ref{rotversusellipticity}).  Following this, the laser power was varied while holding the initial ellipticity and laser detuning fixed (Section~\ref{power}). To investigate the effect of laser detuning, we measured the rotation at a number of probe frequencies ranging over two~GHz and compared these results with theoretical predictions (Section~\ref{rotversusdetuning}).  These measurements were all taken as a function of time.  This permitted us to observe the effects on rotation of the MOT expansion and light-induced atomic motion.

\subsection{\label{preparation}Sample Preparation and Initial Conditions}
\begin{figure}%
\includegraphics[width=\columnwidth]{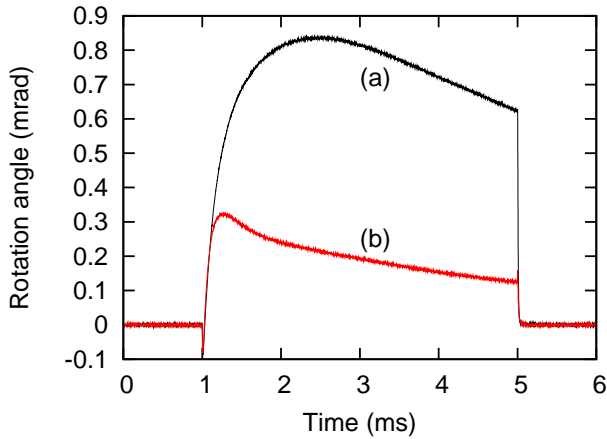}%
\caption{Color online. The probe field self-rotation angle as a function of time, with t = 0 referring to the MOT laser switch off time. We compare the case of the repumper laser on~(a)  and off~(b).  Probe laser power = 600~$\mu$W, detuning = -1~GHz.}%
\label{repump1}%
\end{figure}

\begin{figure}%
\includegraphics[width=\columnwidth]{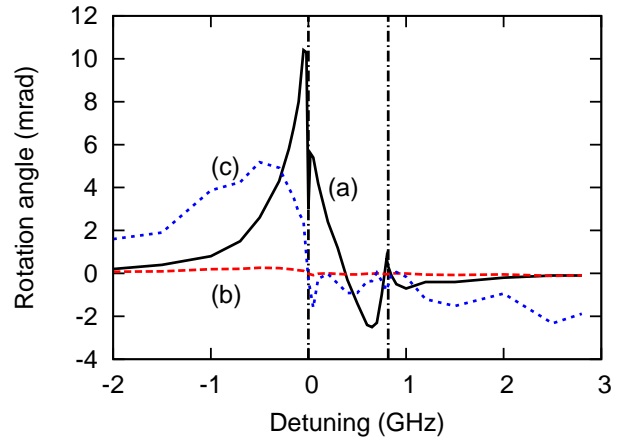}%
\caption{Color online. Rotation angle vs. detuning with repumper laser
on~(a)  and
off~(b). (c) which is the result in (b) but 20 times magnified. Probe laser power =
600~$\mu$W.  Measurements are taken at time 3~ms.
Vertical dash-dot lines mark  locations of the $F_g=2 \to F_e=1$ and 
$F_g=2 \to F_e=2$ D$_1$ line transitions corresponding to 0~GHz and 0.82~GHz detunings.}
\label{repump2}%
\end{figure}

To prevent the influence of the trapping beams on the atoms dynamics, we
turned off the MOT trapping beams while the probe beam was on. In each
40~ms experimental cycle, the trapping beams were turned off (at $t=0$)
for 5~ms, after which the atom cloud recovers during the remaining 35~ms. 
The probe beam was turned on from  $t=1$ to $t=5$~ms. During this 4~ms measurement interval, the atomic cloud expanded due to its thermal motion. The interaction of the atoms with the probe beam resulted in optical pumping into the $F_{_{g}}=1$ ground level. We experimentally observed that, as a result of this effect, if the repumping laser was turned off, a substantially smaller rotation signal resulted.  This effect is illustrated in Fig.~\ref{repump1} comparing the obtained signals with the repumping laser on and off during the measurement interval. In this record, taken at large detuning, leaving the repumping laser on increased the observed rotation  by roughly a factor of three and changed the time evolution of the signal. At smaller detunings, the probe beam has a much greater effect on the atoms. In the absence of the repumper laser, there is almost no rotation (Fig.~\ref{repump2}).  This observation promoted us to leave the repumper on continuously in all other measurements.

We  would  like  to note,  that  although  each  of  the 6  repumper  beams
constituting the  MOT is mostly  circularly polarized, they propagate  in 6
different orthogonal directions  through the cold atom  sample. Further, the
sample  is optically  thin  at the  repumper transition,  so  the light intensity  is
uniform  over the sample volume. For this reason, the optical pumping due to
the repumper generates  unpolarized fluorescence, and thus  does not create
significant polarization  in the ground  states of interest. As  a result, the
repumper does not directly contribute  to the polarization rotation effect
except that it enlarges its strength (as  demonstrated in Fig. 5) because there
are more atoms in the  $F_g=2$ level. In our numerical simulations,
we treat  the presence of  the repumper as  a constant repumping/decay  rate to
sublevels  of  the $5^2S_{1/2}$  $F_g=2$  state  which is  absorbed  into the
$\gamma$ parameter.

The atomic dynamics were also substantially affected by the fact that the atoms are pushed by the radiative force exerted by the probe beam. The corresponding average acceleration is: $a=p \hbar k \Gamma /m$ where $k$ is the light wavenumber, $m$ the atomic mass, $\Gamma$ the excited state radiative decay rate and $p$ the probability for the atom being in the excited state. $p$ depends on the probe intensity and detuning as well as on the repumping rate ($p\leq 1/2$). 
If $p$ approaches 1/2 (a worst case scenario), the acceleration is of
the order $10^{5}$~m/s$^{2}$ for $^{87}$Rb. In the 4~ms interaction
time such acceleration would cause a 0.6 GHz Doppler shift and a 0.8~m
displacement. Although in the experiment the actual value of $p$
is typically much smaller than 1/2 (especially at large detuning) this estimative
gives an indication of how disruptive for the MOT the light pushing
effect could be at small detunings and large probe intensities. 
We attribute spectrally narrow  sharp changes in rotation spectra to the light
pushing effect occurring near the resonance detunings (note them in Fig.~\ref{repump2} of 0 and
0.82~GHz).

The magnetic field gradient necessary for the operation of the MOT is
present in the interaction region.  Attempts to turn off the electric
current in the coils generating the MOT magnetic field resulted in
magnetic transients lasting longer than 10~ms due to eddy currents in the largely
metallic MOT chamber.  Thus, in spite of the anticipated deleterious effect
of this on the purity of the PSR effect, the MOT magnetic field gradient
was left on continuously during the experiment. Although the nature of the
MOT guarantees a zero magnetic field at the center of the atom cloud
through which the probe beam is aligned to pass, the substantial field
gradient means that the atoms, although experiencing a nearly zero average
B field, are nonetheless subject to a spatially inhomogeneous field over the
sample volume.  In addition to this, the radiative forces exerted by the
probe light on the atoms push the atomic cloud away from the zero of the
magnetic field. Since leaving the field gradient on continuously was
necessary to avoid magnetic transients, PSR was studied in a region where
the atoms experience a nonzero magnetic field. Notice that the cloud's thermal expansion and the atomic acceleration induced by the radiative force result in a time varying magnetic environment for the atomic sample.  Such an interpretation is consistent with our observations, as described in the following sections.

\subsection{\label{rotversusellipticity}Rotation vs. Initial Ellipticity}

\begin{figure}%
\includegraphics[width=.8\columnwidth]{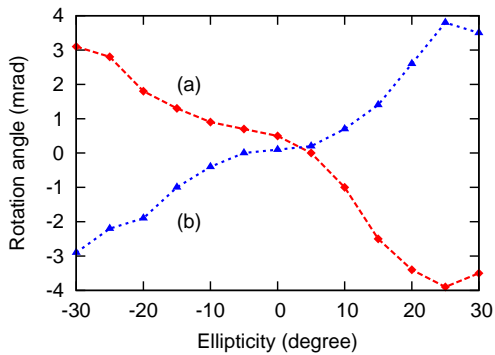}%
\caption{Color online. Probe rotation angle vs. initial ellipticity at 3~ms measured
at 80~MHz~(a) and at -80~MHz~(b) detunings relative to the $F_g=2
\rightarrow F_e=1$ transition.  Probe laser power = 1.8~$\mu$W}%
\label{ellipticity}%
\end{figure}

In order to examine the dependence of the polarization rotation on the probe ellipticity, the incident light ellipticity was varied by rotating a quarter-wave plate placed in the probe beam immediately before it entered the MOT chamber. The ellipticity is given by the angle of rotation of the quarter-wave plate from a reference point corresponding to zero ellipticity.  The measured ellipticity dependence for the probe laser tuned to $-80$~MHz and +80~MHz relative to the $F_g=2 \rightarrow F_e=1$ D$_1$ transition is shown in Fig.~\ref{ellipticity}. As expected, the rotation angle reverses sign around the zero of ellipticity. The small shift of the point of zero rotation with respect to zero ellipticity is a real effect, and is indicative of the presence of a small magnetic field in the measurement region. For the two detunings shown, the rotation has opposite sign as expected from the nearly dispersive shape of the PSR resonances.  These results confirm the occurrence of polarization self-rotation.

\begin{figure}
  \centering
  \subfloat[Probe laser detuning is $-80$~MHz]{\label{ellip color $-80$}
  \includegraphics[width=\columnwidth]{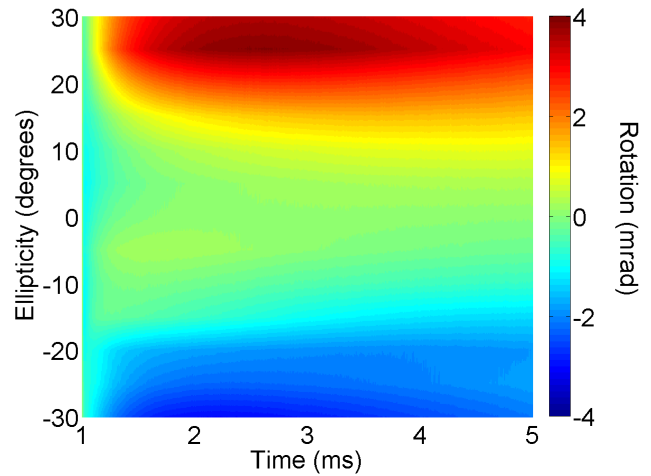}}
  \\
  \subfloat[Probe laser detuning is $+80$~MHz]{\label{ellip color +80}
  \includegraphics[width=\columnwidth]{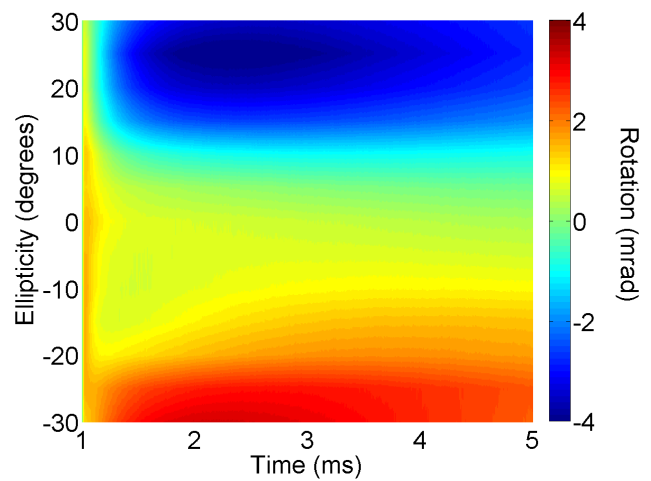}}
  \caption{Color online. Probe laser rotation angle vs. initial ellipticity and time
  measured at two detunings. Probe laser power = 11.4~$\mu$W.}
  \label{ellip color}
\end{figure}

The color map presented in Fig.~\ref{ellip color} shows the time evolution of the polarization rotation for different incident ellipticities at a probe power of 1.8~$\mu$W. A dependence on ellipticity similar to that shown in Fig.~\ref{ellipticity} (at 3~ms) is observed for the entire 4~ms measurement period. A decrease of the rotation is observed for long times; this effect we attribute to MOT expansion.  Given these results, in subsequent measurements of PSR, the probe beam was always given a large initial ellipticity of $\pm 25^\circ$. This ensures that PSR is the dominant process rotating the polarization ellipse of light and selects the largest rotation.

\subsection{\label{power}Rotation vs. Probe Power}

\begin{figure}[h!]
\centering
  \subfloat[Probe ellipticity is $+30^\circ$]{\label{power color +25}
  \includegraphics[width=\columnwidth]{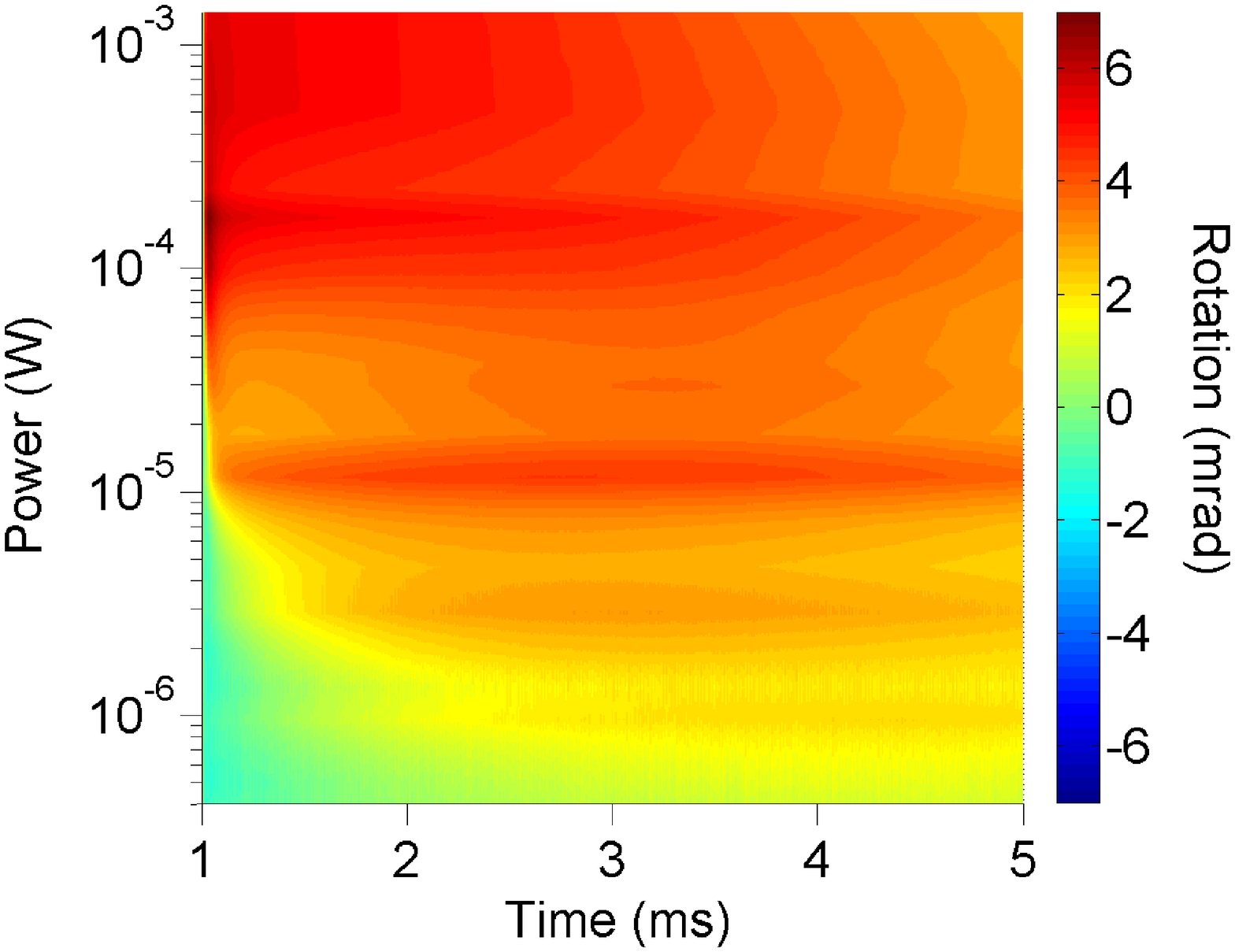}}
  \\
  \subfloat[Probe ellipticity is $-30^\circ$]{\label{power color -25}
  \includegraphics[width=\columnwidth]{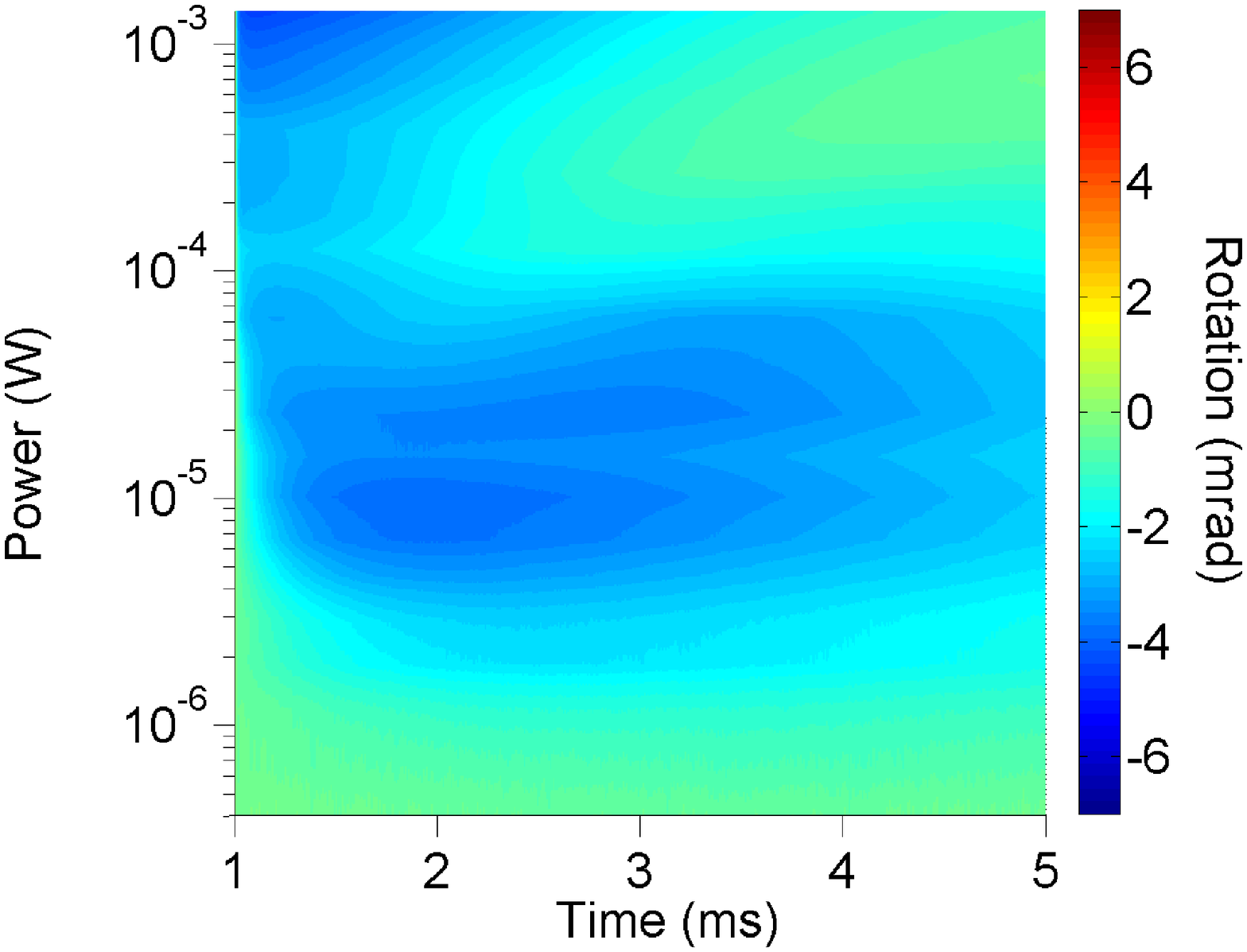}}
  \caption{Color online. Rotation angle vs. probe laser power and time at opposite
  initial ellipticities. Probe laser detuning is $-80$~MHz.}
  \label{power color}
\end{figure}

\begin{figure}[h!]
\centering
\includegraphics[width=\columnwidth]{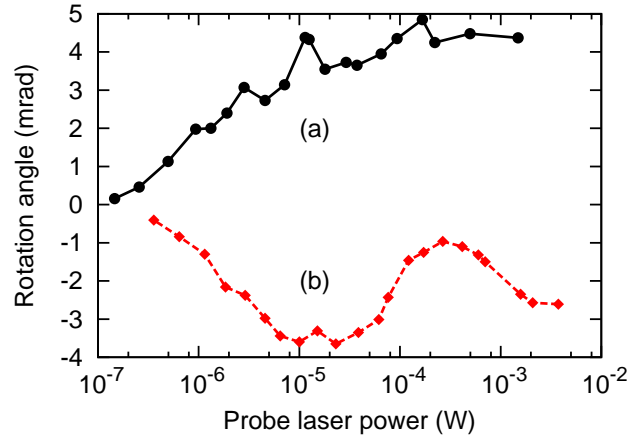}%
\caption{Color online. Rotation angle vs. probe laser power for different
probe ellipticities. Probe detuning is $-80$~MHz, ellipticity
$+30^\circ$~(a) and $-30^\circ$~(b).
}%
\label{power3ms}%
\end{figure}

To investigate the effect of light intensity on self-rotation we measure the rotation angle at various probe powers ranging from 0.3~$\mu$W to 2.0~$mW$. The results are shown in Fig.~\ref{power color} with the probe laser locked at $-80$~MHz from the unperturbed $F_g=2 \rightarrow F_e=1$ transition with both positive and negative incident ellipticity.  A time slice taken at 3~ms expansion time is shown in Fig.~\ref{power3ms}.  Although we expected the self-rotation to increase with laser power, the full dynamics of the observed effect is more complicated.  For low laser powers, the self-rotation effect does appear to steadily increase with power.  However, upon reaching a certain power of the order of 100~$\mu$W,  different behavior is observed depending on the sign of the incident ellipticity. For a positive initial ellipticity, the rotation generally continues to increase with increasing power, but the increase slows and begins to level off.  For the opposite ellipticity, the self-rotation stops increasing and diminishes before increasing again at higher powers.

\begin{figure}
\centering
\includegraphics[width=\columnwidth]{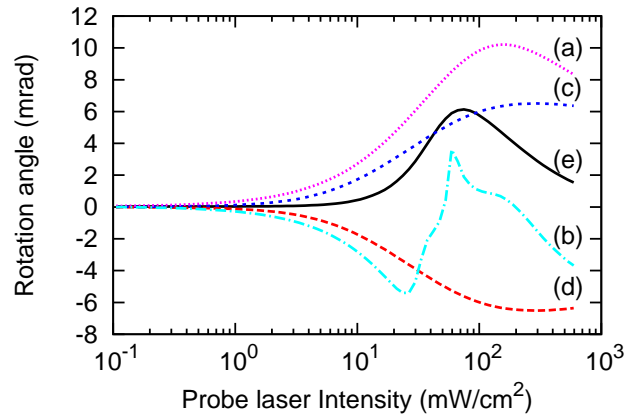}%
\caption{Color online. Simulated rotation angle vs. probe laser power for
different ellipticities and magnetic fields. 
Probe laser detuning is $-80$~MHz, ellipticities are $+30^\circ$~(a,c),
$-30^\circ$~(b,d), and $0^\circ$~(e). Magnetic fields are
B$=0.01\Gamma$~(a,b,e) and B$=0\Gamma$~(c,d).
}%
\label{powerthry}%
\end{figure}

We interpret the different behavior for the two opposite ellipticities as the consequence of the existence of an average nonzero magnetic field in the interaction region. Such a field is present because the probe light pushes
the atoms away from the region of zero magnetic field. As a result of the
MOT symmetry and the initial probe alignment, the mean magnetic field is
oriented along the light propagation axis. In the presence of a magnetic
field, the light polarization experiences a Faraday rotation which has a
nonlinear dependence on light intensity. The Faraday effect becomes
significant as the resonant Rabi frequency of the light becomes comparable
to the detuning. Since the sign of the polarization rotation due to the
Faraday effect is independent of the light ellipticity, its effect enhances
the rotation for one ellipticity and reduces the rotation for the opposite
one. A numerical simulation of the combined PSR and Faraday effect is
presented in Fig.~\ref{powerthry}. The calculation was carried out for a
constant magnetic field $B=0.01 \Gamma$, a figure that corresponds to the estimated field 1 mm away from the MOT center. In the experiment however, as a consequence of the cold atom cloud expansion and the atom acceleration by the probe field, the magnetic environment is variable in time resulting in additional complexity.
The overall explanation for this behavior is twofold.  Depending on the laser detuning, once the probe reaches a certain power, the light begins to have a mechanical effect on the MOT and actually pushes the atoms.  Some atoms can be accelerated out of the MOT while others flow in to take their place, or they can be pushed to areas of nonzero magnetic field, leading to more complicated time dependence of the signal.  We also find that at higher powers, the Faraday effect plays a greater role in the optical rotation leading to what we observe.  Fig.~\ref{powerthry} shows the calculated rotation versus laser intensity, with a small magnetic field included in the calculation, and is seen to display a similar behavior as seen in the experimental results.

\subsection{\label{rotversusdetuning}Rotation vs. Probe Detuning}

\begin{figure}[h!]
  \centering
  \subfloat[Experimental Data taken at 3~ms.]
  {\label{detuning exp}\includegraphics[width=\columnwidth]{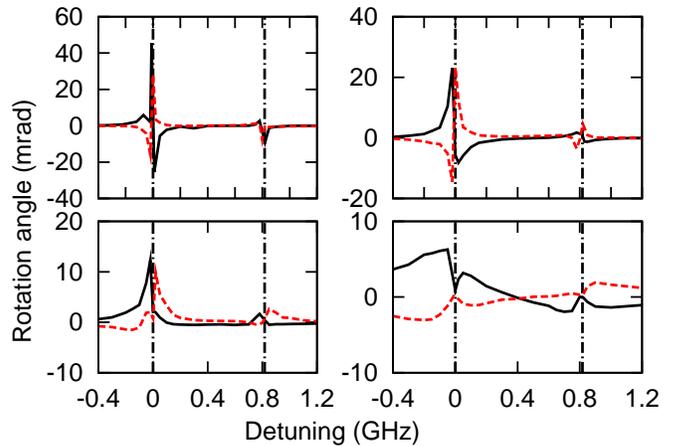}}
  \\
  \subfloat[
    Results of calculations for
    beam cross-section $=10^{-3}$~cm$^2$, B$=0.01\Gamma$, $\gamma = 0.001\Gamma$, and C$=3$.
  ]{\label{detuning thry}\includegraphics[width=\columnwidth]{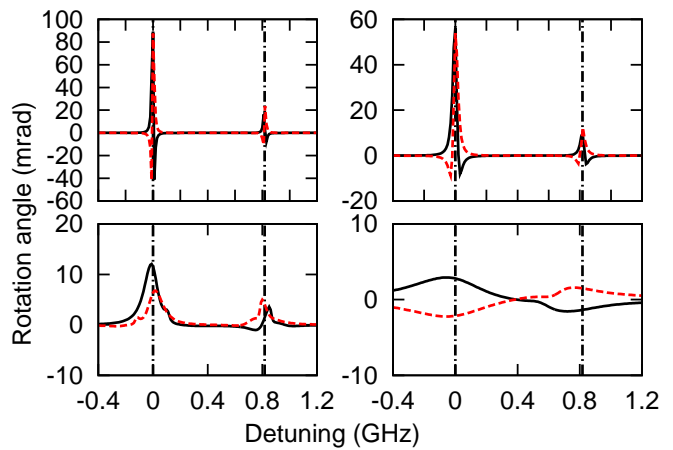}}
  \caption{
    Color online. Comparison of the experimental data and calculated
    rotation angle dependence on  probe laser detuning at opposite initial
    ellipticities $+25^\circ$ (solid lines) and $-25^\circ$ (dashed lines)
    for different probe laser powers: $2~\mu$W (top left), $10~\mu$W (top right), 
    $100~\mu$W (bottom left), and $2000~\mu$W (bottom right).  
    Vertical dash-dot lines mark  locations of the $F_g=2 \to F_e=1$ and 
    $F_g=2 \to F_e=2$ D$_1$ line transitions corresponding to 0~GHz and 0.82~GHz detunings.
    \label{detuning}
  }
\end{figure}

The laser frequency was varied around the $F_g=2\rightarrow F_e=1$ and
$F_g=2\rightarrow F_e=2$ hyperfine transitions during the measurement of the rotation angle in order to obtain the polarization self-rotation spectrum.  The closer the laser frequency is to the transition frequency (small detunings), the stronger the light-atom interaction will be and therefore a larger self-rotation would be expected.  However, in this cold atom system, a smaller detuning also means a stronger mechanical effect of the probe laser accelerating the atoms and disturbing the MOT.  It is clear then that laser detuning and laser power together will determine the self-rotation effects seen in the atomic sample. Fig.~\ref{detuning exp} shows the measured rotation angle versus laser detuning at four different powers and two opposite incident ellipticities.  Fig.~\ref{detuning thry} shows the calculated rotations for similar light intensities assuming the presence of a constant magnetic field along the probe beam propagation direction.  The general trend of the experimental observation is well reproduced by the simulations. As the intensity increases the resonances become power broadened while the peak rotation diminishes. At the highest power the broadening is such that the two hyperfine transitions overlap. For small detunings very close to the resonance ($< 100$~MHz), the self-rotation angle is higher at small powers than for greater laser powers.  This is due to the fact that at higher probe powers and small detunings, the MOT is strongly disturbed due to light pressure and rotation is diminished.  At lower laser powers, the laser frequency can be closer to the resonance while not disturbing the atomic cloud, leading to a relatively higher rotation.  However, at larger detunings, there is almost no rotation in the lower laser power beams, but we do still see some rotation for the higher laser powers.  It appears as though the detuning spectrum for rotation spreads out as the laser power is increased, leading to smaller rotations close to resonance but larger ones at high detunings.  Both transitions must be taken into account, as we can see that the rotation effects from the transitions overlap at high laser powers.

\begin{figure}[h!]
  \centering
  \subfloat[Power 10~$\mu$W,  $\varepsilon=+25^\circ$]{\label{10uW +25}\includegraphics[width=0.5\columnwidth]{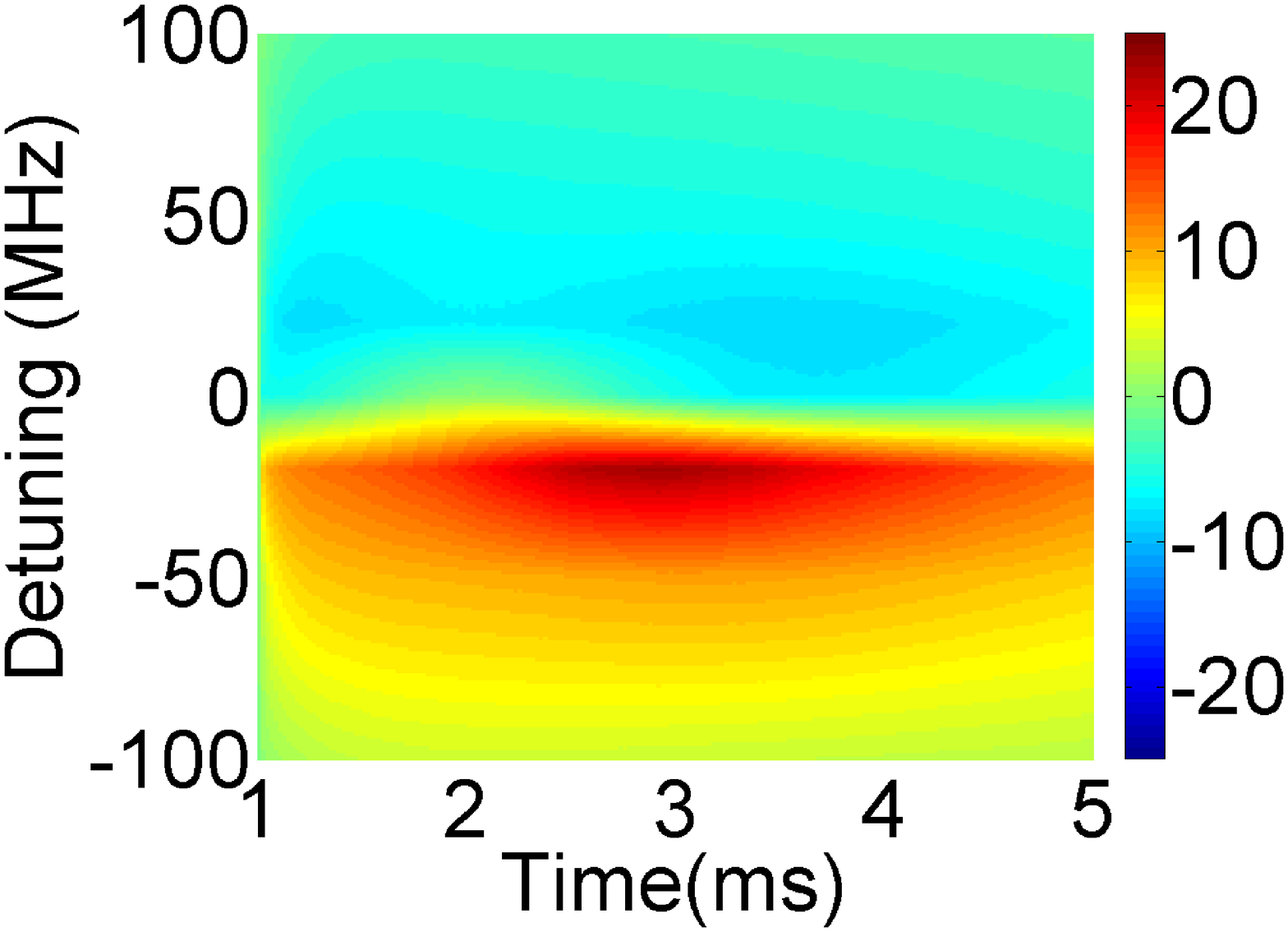}}
  \subfloat[Power 10~$\mu$W,  $\varepsilon=-25^\circ$]{\label{10uW -25}\includegraphics[width=0.5\columnwidth]{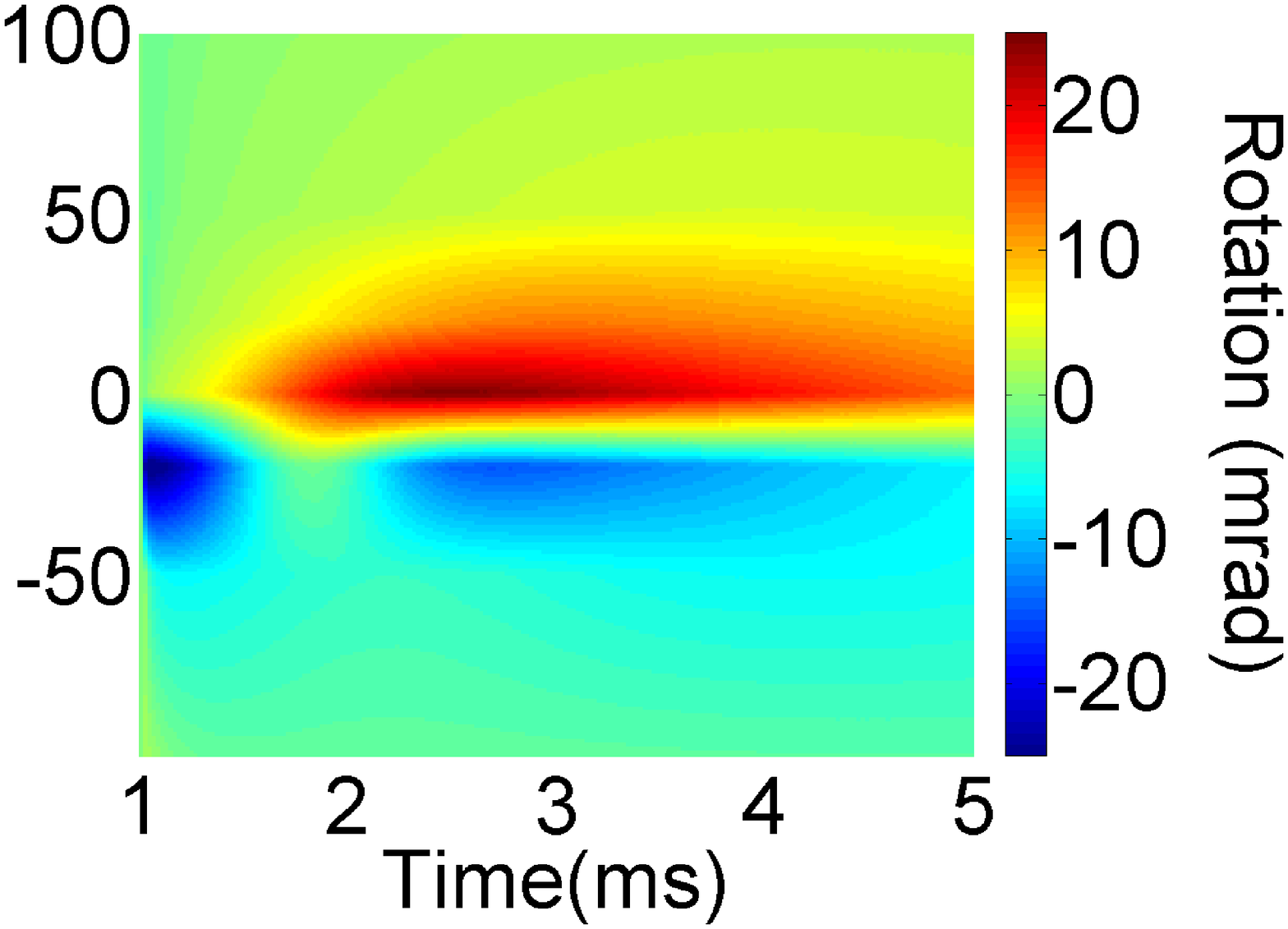}}
  \label{10uW detuning}
\\  \subfloat[Power 600~$\mu$W,  $\varepsilon=+25^\circ$]{\label{600uW +25} \includegraphics [width=0.5\columnwidth]{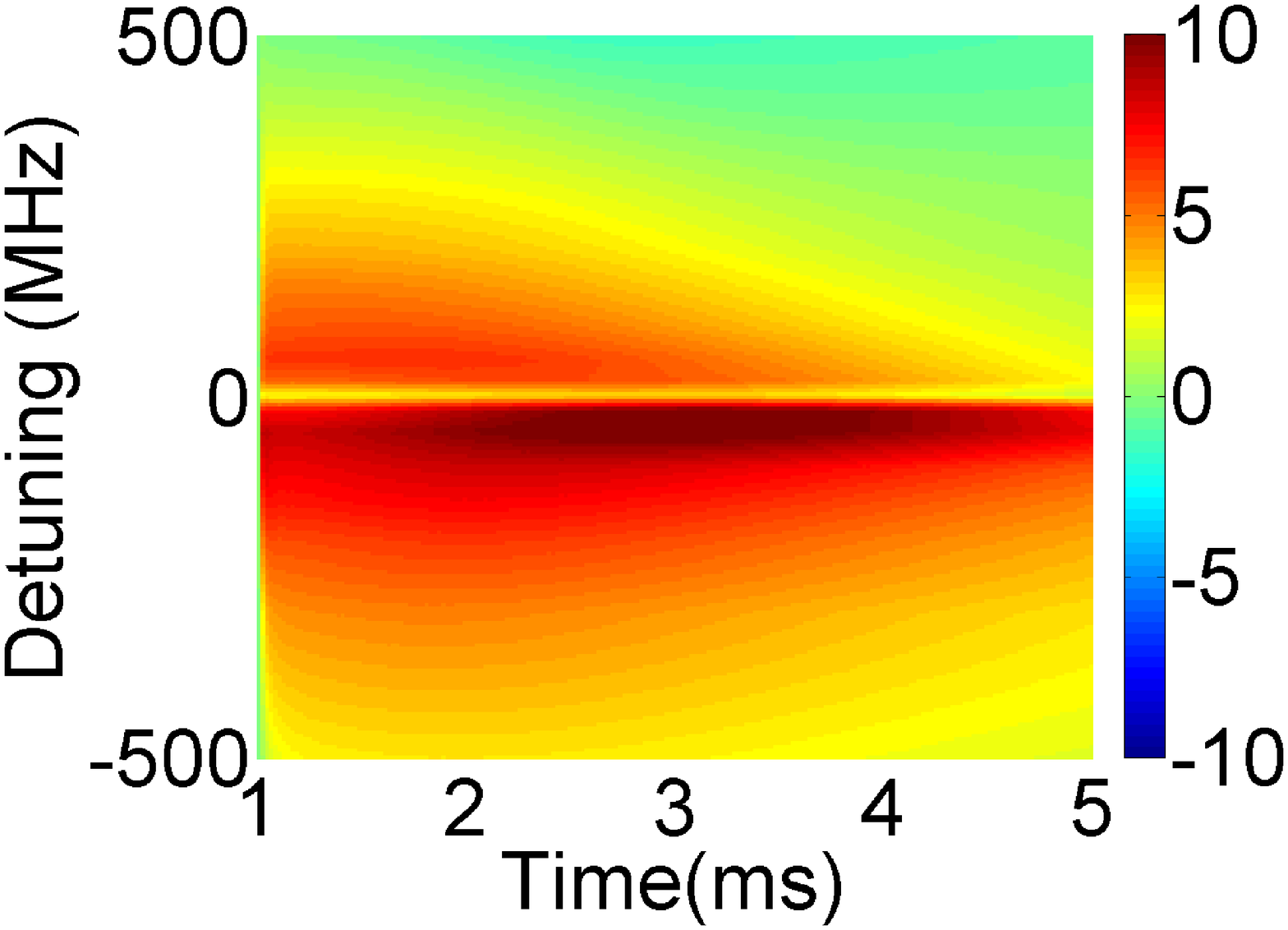}}
  \subfloat[Power 600~$\mu$W,  $\varepsilon=-25^\circ$]{\label{600uW -25}\includegraphics [width=0.5\columnwidth]{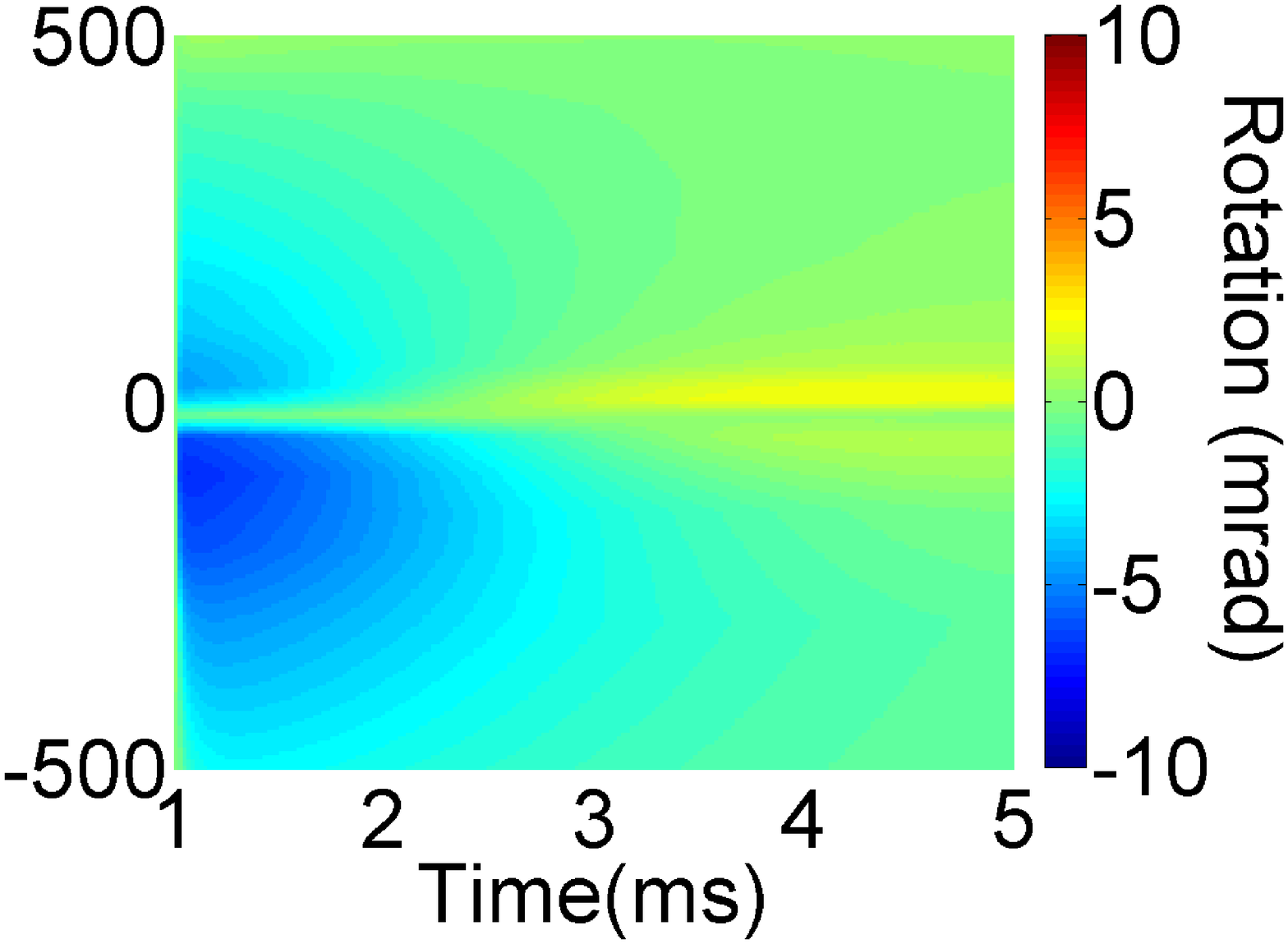}}
  \caption{Color online. Dependence of rotation angle on probe laser  detuning and
  measurement time for different probe laser powers and ellipticities.
  \label{detuning color}
  }
\end{figure}

This general complex interdependence of the self-rotation angle on laser
detuning and different powers is qualitatively plausible.  However, beyond
that, we see that the experimental data matches the calculations fairly
well in shape and in size of the self-rotation angle. The main difference
between the experimental plots and the simulations is that in the
experiment, the rotation is always nearly zero at zero detuning, but not in
the simulations. 
We believe that this is due to the strong perturbation of
the atom cloud caused, near resonance, by the radiative force produced by
the probe beam (compare bottom rows of
Fig.~\ref{detuning exp} and Fig.~\ref{detuning thry} at near resonance
detunings).
Such a mechanical effect is not accounted for in the
simulations. The overall agreement between the simulation and the observed
spectra strongly supports the assumption of an average nonzero magnetic
field throughout the sample. Because the Faraday rotation is highest at
zero detuning, we see high rotation here in the simulations, especially at higher powers when the Faraday effect dominates over PSR.  The similarity between the experimental data and the calculated data is fairly good at larger detunings where the MOT is not disturbed.  We also note that the asymmetry between positive and negative rotations and the difference between the strength of the first and second transitions.  The overall agreement between measurements and the simulations strongly support the conclusion that we have a non-zero magnetic field in the experiment.  We see effects of this field because parts of the atomic cloud are in regions of non-zero field as it expands, or because the light pressure pushes atoms to a region of nonzero field. The experimental data shown in Fig.~\ref{detuning exp} was taken at 3~ms of cloud expansion. The complete time evolution is shown in Fig.~\ref{detuning color}. The time dependence is understandably stronger for smaller detunings where the light has a stronger mechanical effect, pushing the atoms and disturbing the MOT.

\section{Summary and Outlook}
We have studied several aspects of polarization self-rotation in cold rubidium atoms.  We have focused our study on the $F_g=2\rightarrow F_e=1$ and $F_g=2\rightarrow F_e=2$ D$_1$ hyperfine transitions of $^{87}$Rb.  We find that with this experimental setup, PSR is readily observable. As expected, the rotation depends on the incident ellipticity of the light. The rotation depends on the probe power, growing with increased power at large detunings.  However, at higher laser power, the probe beam begins to disturb the MOT, pushing atoms away from the trapping zone and to a region where the magnetic field is no longer zero, on average. The rotation at these higher powers does not appear to continue increasing with power, but becomes less quantitatively predictable due to the motion of the atoms and the non-zero magnetic field.  We see from the rotation measurements at different detunings that the effect we observe is not symmetric around the transition confirming the fact that many of the atoms contributing to self-rotation are experiencing some small magnetic field. If a longitudinal static magnetic field is included in the self-rotation numerical simulations, the measured behavior is quite similar to the calculations which include the effect of Faraday rotation. We also point out that, with increasing laser power, the rotation spectrum in frequency space appears to spread and the rotation decreases due to the increased light pressure the probe has on the atoms as well as a greater magnetic field effect. We can observe self-rotation at very large detunings upwards of 500~MHz at higher laser powers approaching milliwatt levels.

Although PSR in cold atoms may be a useful tool for testing and monitoring
the magnetic field environment affecting the atomic cloud in a MOT chamber,
there are a number of other techniques that are already in use to this end.
The main emphasis here is observation of new effects due to the interplay
of PSR and the usual Faraday effect.  Polarization self-rotation is also
promising for generation of squeezed vacuum states.  We have in fact
observed indications of PSR-based squeezing in ultracold samples; these results will be presented in a forthcoming paper.

Finally, we point out that it is likely that MOT characterization and squeezed light generation applications will require larger PSR angles than those presented in this paper.  The main restriction in this experiment, preventing the detection of higher self-rotation angles, comes from the limited optical depth of the atomic sample.  With our beam size and MOT atomic density, the probe interacts with only about $10^5$ atoms resulting in the above mentioned optical depth of $\sim 2$.  However, in cell experiments, where substantial squeezing has been observed, the probe laser interacts with about a thousand times more atoms and so higher self-rotations are obtained.  A solution to this problem would be to create a MOT with a higher optical depth.  It is possible with current technology to create MOTs with up to $10^{10}$ atoms, much high densities~\cite{1402-4896-81-2-025301}, and optical depths on the order of several hundred.  It is also possible to create atomic clouds with different geometries~\cite{Greenberg:07, Lin:08} which could be quite beneficial for PSR studies in ultracold atomic physics.  For example a cigar-shaped MOT with the probe aligned along its major axis would give a much larger interaction length and therefore larger optical depth.

\begin{acknowledgments}
E.M. and T.H. thank the support of NSF Grant PHY-0758010. 
A.L. wishes to thank support from ANII, CSIC, PEDECIBA
(Uruguayan agencies), and the APS International Travel Grant Program.
Partial support of this work (S.B and M.H.) was provided by NSF Grant PHY-0654226.
Numerical simulations were carried in the SciClone Cluster (College of
William and Mary). 
\end{acknowledgments}


\bibliographystyle{apsrev}

\end{document}